\pdfoutput=1
\documentclass[preprintnumbers,endnote,nofootinbib,preprint]{revtex4}
\usepackage{graphicx}
\usepackage{amsmath,array}
\usepackage{multirow}

\newcommand{\lsim}{\lesssim}
\newcommand{\gsim}{\gtrsim}

\newcommand{\ord}[1]{\mathcal{O}{(#1)}}
\newcommand{\beq}{\begin{equation}}
\newcommand{\eeq}{\end{equation}}
\newcommand{\bea}{\begin{eqnarray}}
\newcommand{\eea}{\end{eqnarray}}

\newcommand{\mP}{\bar{M}_{\rm P}}

\newcommand{\invfb}{${\rm fb}^{-1}$}

\def\misse{E\hspace{-0.2cm}/_{T}}

\usepackage{color}

\begin{document}
\pagestyle{plain}
\title{\boldmath Photon Cascade Decay of the Warped Graviton at LHC14 and  a 100 TeV Hadron Collider
%
%
}
\begin{flushright}
%
%
UMD-PP-014-027 \\
\end{flushright}
\author{Kaustubh Agashe$^{\, a}$
}
\author{Chien-Yi Chen$^{\, b}$
}
\author{Hooman Davoudiasl$^{\, b}$
}
\author{Doojin Kim$^{\, c}$
}
\affiliation{
\vspace*{.5cm}
 \mbox{$^a$Maryland Center for Fundamental Physics}, \\
 \mbox{University of Maryland, College Park, MD 20742, USA}\\
 \mbox{$^b$Department of Physics, Brookhaven National Laboratory, Upton, NY 11973, USA}\\
 \mbox{$^c$Department of Physics, University of Florida, Gainesville, FL 32611, USA}\\
 \vspace*{1cm}}
 
\begin{abstract}

In warped 5D models of hierarchy and flavor, the first Kaluza-Klein (KK) state of
the graviton $G_1$ is heavy enough to decay into a photon and its first
KK mode $\gamma_1$ on-shell: $G_1 \to \gamma_1 \gamma$.    
The volume-suppression of the rate for this process [relative to 2-body decay into heavy Standard Model (SM) final states ($W/Z/t/H$)] 
may be partially compensated by the simplicity of the photon final state.
We consider $\gamma_1 \to W^+W^-$, with a typical $\ord{1}$
branching fraction, and focus on the semi-leptonic final state $W(\to jj) W(\to \ell, \nu)$ with $\ell=e,\mu$. 
The SM background originates from $2\to 3$ parton processes and is relatively suppressed compared to those for 2-body decays of $G_1$.
Moreover, to further reduce the background, we can impose an
%
%
invariant mass window cut
%
%
for $\gamma_1$ (in addition to that for $G_1$) in this new channel.
We emphasize that 
this ``photon cascade" decay probes a {\em different} combination of 
(bulk and brane) interactions of the KK states than the decays into two heavy SM states. 
Thus, in combination with other channels,
the cascade decay could be used to extract
the {\em individual} underlying geometric parameters.  
The $3\sigma$ reach for $G_1$ in our channel is up to 1.5~TeV at the high luminosity (14 TeV) LHC, 
and can be extended to about 4~TeV, at $5\sigma$, at a future 100~TeV hadron collider.
Along the way, we 
point out the novel
%
%
feature
that the 
invariant mass distribution of KK graviton decay products becomes skewed from the Breit-Wigner form, 
due to the KK graviton coupling growing with energy.

\end{abstract}
\maketitle


\section{Introduction}

The hierarchy between the Higgs
boson mass $m_H \simeq 126$~GeV \cite{Aad:2012tfa,Chatrchyan:2012ufa}
and high scales can be addressed by the Randall-Sundrum (RS)
model \cite{Randall:1999ee}.  This model is based on a warped background geometry that is a slice of AdS$_5$,
with an associated curvature scale $k$.  The 5D RS background is bounded by 4D UV (Planck) and IR (TeV) branes.
In this background, UV brane scales, typically of order $\mP \approx 2 \times 10^{18}$~GeV (i.e., the Planck scale),
get warped down along the fifth dimension and become $\ord{1~\text{TeV}}$ at the IR brane: $\mP \, e^{-k\pi R}\sim$~TeV, for
$k R \approx 11$, where the size of the compact fifth dimension is $\pi R$.
Thus, the hierarchy between the Planck scale and the weak scale (or Higgs boson mass) is exponentially generated by using moderate-sized underlying parameters.

The original RS model assumed that the {\em entire} Standard Model (SM) was confined to the IR brane \cite{Randall:1999ee}.
The signals of this setup were thus only from the gravitational
sector \cite{GW1,Davoudiasl:1999jd,GW2}, including a characteristic tower of spin-2 Kaluza-Klein (KK)
gravitons, whose masses and interactions with {\em all} SM particles are governed by the TeV-scale.
This leads to distinct dilepton or diphoton collider signals of the
TeV-scale spin-2 resonances (via their production from both quarks and gluons in the initial
state) \cite{Davoudiasl:1999jd}.

In the RS proposal, the Higgs field needs to be localized close to the IR brane in order to explain
the Planck-weak hierarchy, however other fields can
propagate in the 5D bulk \cite{Goldberger:1999wh}.  It was realized that by allowing
the SM gauge fields \cite{Davoudiasl:1999tf,Pomarol:1999ad} and fermions \cite{Grossman:1999ra}
to propagate in the bulk, one can obtain a model of flavor \cite{Gherghetta:2000qt, Huber:2000ie, Huber:2003tu}, as well as Planck-weak hierarchy.
Here, 5D masses for fermions \cite{Grossman:1999ra} control their zero-mode
profiles and can lead to a wide range of 4D Yukawa couplings with natural values for the bulk parameters.
Thus, this 
framework is very attractive, providing solutions to both the Planck-weak and flavor hierarchy problems 
of the SM.

The warped models of flavor and Planck-weak hierarchies have additional signatures
associated with the KK states of the gauge and the fermion fields.  However, all these
KK modes (including that of graviton) interact mostly with heaviest SM fields
because all the particles involved in these couplings are localized near the TeV/IR brane; the KK modes 
have suppressed couplings to other SM particles which are more spread in the bulk.
Thus, the KK modes decay, once produced, dominantly into $t$ or Higgs (including {\em longitudinal} $W/Z$).
In particular, 
the KK graviton couplings to light fermions are
quite tiny, due to the latter's profiles being localized near the Planck brane.  
Because of profiles of massless photons and
gluons being flat, the couplings of KK gravitons to them are also small, but not negligible, being suppressed by a ``volume factor" $\sim 1/(k\pi R)$ (compared to $\sim 1 / \hbox{TeV}$) \cite{Davoudiasl:2000wi}.
So, it is these KK graviton couplings to {\em gluons} (i.e., not to light quarks) which 
have to be used for their production (and again, decays are dominantly to top quark/$W$/$Z$/Higgs).
Hence, overall, graviton collider phenomenology is significantly different from that of the original RS model,
and the associated signals are more elusive and challenging to reconstruct.

Collider phenomenology of warped models of hierarchy and flavor (bulk SM) have been studied, for a example, in
Refs.~\cite{Davoudiasl:2000wi, Agashe:2006hk,Lillie:2007yh,Fitzpatrick:2007qr,Agashe:2007zd,Lillie:2007ve,Djouadi:2007eg,
Agashe:2007ki,Antipin:2007pi,Agashe:2008jb,
Davoudiasl:2009cd}.  Specifically, the first KK graviton $G_1$ signals at the LHC have been examined, using $G_1\to t{\bar t}$ \cite{Fitzpatrick:2007qr},
$G_1\to Z_L (\to \ell^+\ell^-)Z_L(\to \ell^+\ell^-$), $\ell=e,\mu$ (with $Z_L$ denoting longitudinal $Z$) \cite{Agashe:2007zd}, $G_1\to W_LW_L$ \cite{Antipin:2007pi}, and $G_1 \to Z_L(\to \ell^+\ell^-)Z_L(\to \nu \bar\nu)$ \cite{Chen:2014oha}.
The results of these studies suggest that $G_1$ up to $\sim$ 2~TeV with
$\sim {\rm few}\times 100$~\invfb of data can be discovered at the 14~TeV LHC.\footnote{In Ref.~\cite{Agashe:2007zd}, a factor of 1/2 is left out in the
matrix element for $gg\rightarrow G_1 \to VV$ and hence the associated cross section should be smaller by 1/4.  The reach of the LHC for KK gravitons in Ref.~\cite{Agashe:2007zd} is therefore less than the projected $\sim$ 2 TeV.}
For other studies of RS models at future colliders, see also Refs.~\cite{Agashe:2013fda}.

In fact, the ATLAS and CMS collaborations have already actively searched for these KK particles (including gravitons) at the LHC7/8: see references \cite{Khachatryan:2014gha, ATLAS:039, Chatrchyan:2013lca,TheATLAScollaboration:2013kha}.
The focus here was mostly on KK particle decays into 2 SM particles,
and the bounds are already in the $\sim 1$ TeV range (depending on the KK particle and the 
exact model parameters).
Given this absence of any signal thus far, it is possible that 
even the KK particles' discovery at LHC14 might be difficult, let alone measurements of their properties.

At the same time, the possibility of a 100 TeV hadron collider is being seriously considered \cite{workshop}.
Such a collider 
could afford not just discovery of the KK particles, but also allow their precision studies.
In the light of this situation, it seems timely and interesting to study the KK particle 
signals in more detail, including their other decay modes, especially if these can 
yield additional information about model parameters.  
With this goal in mind,  in this work, we examine the prospects for detecting novel signals of KK {\em gravitons} in their cascade decays, i.e.,
those with final states containing other KK particles, at hadron colliders.  In particular, we
focus on
\beq
pp \to G_1 \to \gamma_1 \gamma,
\label{prod}
\eeq
where $\gamma_1$ denotes the first KK mode of the photon $\gamma$.
Note that the masses of KK gravitons in warped models are given by
$m^G_n = x^G_n \, k e^{-k \pi R}$, with $x^G_n = 3.83, 7.02,\ldots$, whereas for
the masses of gauge boson KK modes we have $m^A_n = x^A_n \, k e^{-k \pi R}$,
with $x^A_n=2.45, 5.57,\ldots$ ($n=1,2,\dots$).
Hence, the above two-body decay can occur on-shell, with sizeable phase-space.
While such
a study is novel for the KK {\em graviton} (to our knowledge, done for the first time here), 
cascade decays of other KK particles have been considered before, 
for example, KK gluon $\rightarrow$ top $+$ KK top or 2 KK tops: 
see references \cite{Carena:2007tn,Bini:2011zb,Barcelo:2011wu,Greco:2014aza}. However, the availability of such a phase space 
might not be as robust for KK gluon as it is for KK graviton (as mentioned above).

A brief comparison of this new channel with those containing SM conjugate pairs is in order here.
It is true that the $G_1\gamma_1\gamma$ coupling is suppressed by a
factor $\sim 1/\sqrt{k\pi R}$ (as compared to decays to 2 heavy SM states), due to the flat profile of 
the photon. However, the detection of the photon, unlike heavy SM states (such as $W/Z$/top/Higgs),
does not require reconstruction.  This is a good feature, since reconstruction of $W/Z$/top/Higgs
(especially given their large boost in this case, in turn, due to KK graviton being much heavier) 
often leads to reduced signal strength due to low efficiency or the
small branching fractions of clean final states.  We will focus on the case
$\gamma_1 \to W^+ W^-$, which can typically have $\ord{1}$ branching
fraction \cite{Agashe:2007ki}, and consider the final state corresponding to $WW\to jj \ell \nu$.  
The relevant irreducible background here
is $W^+W^-\gamma$, which is a $2\to 3$ process at the parton level.  This
background is thus relatively suppressed by phase space compared to
$2\to 2$ backgrounds for $G_1$ decaying into a pair of SM states,
such as $W^+W^-$ [however, as we will see, the dominant background in our analysis is from $Wj\gamma$, due 
to the large boost of $W(\to jj)$ which makes it appear as one jet].  
Also, in principle, one can impose {\em two} invariant mass window cuts (for KK graviton and KK $\gamma$) vs. only one for decays into
2 SM particles.

These observations imply that $G_1\to \gamma_1 \gamma$, while having a
small production cross section, can entail certain advantages regarding its signal to background
ratio compared with other more conventional channels.  
Thus it could potentially be a good cross-check 
to the 2 heavy SM final states as a probe of the KK graviton.
Finally, in the post-discovery and precision study phase, the specific dependence of the rate on the
volume factor for the process (\ref{prod}), which is different from the conventional ones,   
in conjunction with information from other channels, may lead to the extraction of 
the underlying model parameters {\em separately}.
So, we emphasize that, in general,
this new process is complementary to the ones previously studied.

Here is an outline of the rest of this paper. We begin with a brief review of the model, in particular, the couplings of KK graviton 
as relevant for its 
production and decay.
In Section \ref{cuts}, we delineate
the set of 
cuts which are employed in order to enhance the KK graviton signal-over-background.
The resulting discovery reach for KK graviton is presented 
in Section \ref{results}.
We conclude in Section \ref{conclude}.

\section{The Model}
\label{model}


We will summarize the basics of the warped extra-dimensional framework with bulk SM that we are studying here: for more details, 
see \cite{Davoudiasl:2009cd}.

\subsection{Basics of profiles, masses and couplings}

The metric is given by
\begin{eqnarray}\label{eq:yzcoordinate}
ds^2 & = 
& \frac{1}{ ( k z )^2 } \left(  \eta^{ \mu \nu } dx_{ \mu } dx_{ \nu } - dz^2 \right)
%
%
\end{eqnarray}
where $k$ is AdS curvature scale and 
%
%
$z = 1/k$ and $\exp ( - k \pi R ) / k$ 
correspond to the UV and IR branes (often called Planck and TeV branes), respectively. $R$ denotes the radius of the extra-dimension. The KK masses are quantized in units of $k \exp \left( - k \pi R \right)$, which is taken to be of TeV-size.
The SM gauge and fermion fields propagate in the extra dimension with the Higgs field being assumed to be (eaxctly) localized on the TeV brane (for simplicity).
In the (effective) 4D theory, these 5D fields appear as towers of modes via the standard KK decomposition: zero modes correspond
to SM particles and the heavier modes are denoted as KK particles.
This procedure also gives these masses and profiles of  4D modes. 
The couplings of this model depend on the overlap in the extra dimension of the profiles of the particles involved.


As mentioned in the introduction, a (rough) sketch of the profiles is as follows:
\begin{itemize}

\item
{\em all} KK modes (like the Higgs) are localized near the TeV/IR brane

\item
top/bottom quark are either localized near TeV brane or have a roughly flat profile (based on requirement of their large coupling to Higgs)

\item
SM photon and gluon have flat profiles

\item
light SM fermion profiles are peaked near the Planck/UV brane, in accord with their small coupling to Higgs.

\end{itemize}

Here we focus on production at the LHC (and 100 TeV collider) of KK graviton, followed by its decay into a KK photon and photon.
However, for this purpose, we also need to consider the {\em total} decay width of the KK graviton, in particular, the (other) dominant decay
modes of the KK graviton.
So, the relevant features of the KK graviton couplings (based on overlaps of above profiles) are:
\begin{itemize}

\item
largest coupling is to top/bottom quarks; Higgs (including longitudinal $W/Z$);

\item
coupling to SM gluon, as used in production, is suppressed by volume, i.e., logarithm of Planck-weak hierarchy (similarly
for photon and transverse $W/Z$) and

\item
coupling to desired final state, i.e., KK photon, plus photon is ``in-between" above two.

\end{itemize}

\subsection{Details of choice of KK masses}

We now discuss what would be a reasonable choice for values of KK masses for such a study, based on the rough guide of naturalness  
(i.e., TeV-ish), combined with current bounds.
The point is that the interactions of the SM fields with the KK modes in  warped hierarchy/flavor 
models can lead to various deviations from precision electroweak (EW) and flavor data.  Hence, to avoid 
finely tuned parameters, one often needs to introduce various new symmetries that would 
control the size of such deviations.  In particular,  custodial isospin \cite{Agashe:2003zs} allows gauge KK 
masses as small as $m_{KK}^{gauge}\sim 3$ TeV  to be consistent 
with oblique EW data, and extension of these symmetries \cite{Agashe:2006at} would also suppress deviations in 
$Z b \bar{b}$ coupling sufficiently for such KK masses  \cite{Carena:2006bn,Carena:2007ua, Delaunay:2010dw}.  
%
%
%
%
%
%
As far as consistency with flavor changing neutral currents (FCNC's) and CP-violating processes is concerned, in spite of a built-in analog of GIM mechanism of the SM 
\cite{Gherghetta:2000qt, Huber:2000ie, Agashe:2004cp}, a KK mass scale of at least 
$\sim 10$ TeV seems necessary \cite{Csaki:2008zd}. However, upon suitable augmentation by various 
approximate flavor symmetries, 
$m_{KK}^{gauge}\sim 3$ TeV can be consistent with these bounds (see \cite{Barbieri:2012tu}
%
%
for recent work in a ``simplified" version of the 5D model).

The above discussion indicates that based on {\em naturalness} arguments alone, one would expect 
generic warped KK states to emerge above $\sim 3$~TeV or so.  However, as mentioned earlier, direct LHC bounds 
on KK gravitons of interest in our work are at or below $\sim 1$~TeV.  Given the diverse set of choices that model-building 
can provide, we will hence focus our attention on the direct collider bounds and consider values of KK masses that may 
otherwise be disfavored in particular models based on fine-tuning arguments.

\subsection{Couplings to dominant decay modes and total decay width of KK graviton}

As mentioned above, the decays of
KK graviton are dominated by top
quark and Higgs (including longitudinal $W/Z,$
using Goldstone equivalence theorem).  Let us consider the
top and bottom sector in more detail to determine
their couplings to KK graviton. 
With a suitable custodial symmetry to
relax the $Zb\bar b$ constraint, various configurations 
are possible: (i) $t_R$ very close to TeV brane with $(t,b)_L$ having
a profile close to flat \cite{Agashe:2003zs}, (ii) $(t,b)_L$ very close to the TeV brane
and $t_R$ close to flat,
and also (iii) the intermediate possibility with both
$t_R$ and $(t,b)_L$ being near, but not too close to
TeV brane.

For simplicity, in our actual analysis, we will consider the 
case with $t_R$ 
localized very close to the
TeV brane, with $(t,b)_L$ having close to a flat profile.  
It is straightforward to extend our analysis
to the other cases (as we will briefly indicate below).  
Moreover, we will assume that
this helicity of the top quark and similarly the Higgs
are {\em exactly} localized on the TeV brane.  
In reality, these particles have a profile {\em peaked}
near the TeV brane, but this will result in
at most an $O(1)$ difference.

With our assumptions and approximations above, the couplings relevant for calculation of the {\em total} decay width
of the KK graviton are simply:
\begin{eqnarray}
{\cal L}_G & \ni & \frac{e^{ k \pi R } }{ \mP }
\eta^{ \mu \alpha } \eta^{ \nu \beta } h^{ ( q ) }_{ \alpha \beta }
( x ) T_{ \mu \nu }^{ t_R, H } ( x )
\label{G1tRH}
\end{eqnarray}
giving the partial decay widths \cite{Han:1998sg}:
\begin{eqnarray}
\label{Gtott}
\Gamma \left( G \rightarrow
t_R \bar{ t_R } \right) & \approx & N_c \frac{ (c\, x^G_n)^2 \, m^G_n }
{ 320 \pi } \\
\Gamma \left( G \rightarrow H H \right) & \approx &
 \frac{ (c\, x^G_n)^2 \, m^G_n }
{ 960  \pi } \\
\Gamma \left( G \rightarrow W^+_L W^-_L \right) & \approx &
 \frac{ (c\, x^G_n)^2 \, m^G_n }
{ 480  \pi } \\
\Gamma \left( G \rightarrow Z_L Z_L  \right) & \approx &
 \frac{ (c\, x^G_n)^2 \, m^G_n }
{ 960  \pi }
\end{eqnarray}
where $N_c = 3$ is the number of QCD colors, $c\equiv k/\mP$,
and we have neglected
masses of final state particles
in phase space factors.  These are the only important decay
channels entering the total width for the $n=1$ graviton KK mode which is the focus of
our analysis in this work.
The last 2 formulae correspond to decays to longitudinal
polarizations: we have used
equivalence theorem
(which is valid
up to $M_{ W, Z }^2 / E^2$ effects, where $E \sim m^G_1$) to relate
these decays to physical Higgs.
As mentioned above, we can
neglect decays to transverse $W/Z$ (and similarly to gluon, photon)
due to volume
[$\sim \log \left( \mP
/ \hbox{TeV} \right)$] suppression (in amplitude)
relative to longitudinal polarization. 
Similarly, 
$\left( t,b \right)_L$ can be dropped here, due to their close-to-flat profiles.
And, of course, 
decays
to light fermions are completely negligible (due to the Yukawa-suppressed
coupling to KK graviton).
We can also show that the decays of KK graviton
to other KK modes -- such as in our channel -- are suppressed and hence 
can be neglected {\em as far as the total width calculation is concerned}.
Thus, total width of the KK graviton is given (approximately) by
\begin{eqnarray}
\Gamma \left( G \rightarrow \hbox{all} \right) & \approx & 
 \frac{ 13 (c\, x^G_n)^2 \, m^G_n }
{ 960  \pi } \label{eq:G1decawidth}
\end{eqnarray}

We would like to emphasize here that,
in general, the total decay width should be taken as a free parameter.  For example, ``switching" top quark LH vs.~RH profiles gives
a total width of $\Big[  22 \left(  c\, x^G_n \right) ^2 \, m^G_n \Big] /
\left(  960  \pi \right)$.
The reason is 
that for the case where
$(t,b)_L$ is localized very close to the
TeV brane (with $t_R$ being close to flat),
we multiply the right hand side of Eq.~(\ref{Gtott}) by a factor of $2$ to include decays
to $b_L$.
In this case,
production of KK graviton from $b \bar{b}$ annihilation
can also be important. 
On other hand, for the intermediate possibility mentioned above
[with both
$t_R$ and $(t,b)_L$ being near, but not too close to
TeV brane, unlike the Higgs], the partial width of KK graviton
to top/bottom quarks (and hence the total width) will be smaller, given roughly by 
$\Big[  4 \left( c\, x^G_n \right)^2 \, m^G_n \Big]  /
\left(  960  \pi \right)$. 
Hence the
BR to our desired final state will be larger (albeit still small).
Finally, there is the possibility that the KK graviton decays into {\em additional} light states localized near the TeV brane  (such as the radion), which can increase the total decay width.

\subsection{Details of KK graviton coupling to initial and final states of interest}


%
The couplings of KK graviton to the initial state (gluons) and desired final state (KK photon plus photon)
are somewhat more involved than the ones given above (which dominate the decay width).
A schematic formula for couplings of $m^{ \hbox{th} }$ and
$n^{ \hbox{th} }$ modes of the bulk field (denoted by $F$)
to the $q^{ \hbox{th} }$ level KK gravitons (denoted by $G$)
is \cite{Davoudiasl:2000wi}:
\begin{eqnarray}
{\cal L}_G & = & \sum_{ m , n, q } C^{ F F G }_{ m n q } \frac{1}{ \mP }
\eta^{ \mu \alpha } \eta^{ \nu \beta } h^{ ( q ) }_{ \alpha \beta }
( x ) T_{ \mu \nu }^{ ( m, n ) } ( x )
\label{general}
\end{eqnarray}
where $h^{ (q) }_{ \alpha \beta }
( x )$ corresponds to the KK graviton, $T_{ \mu \nu }^{ ( m, n ) } ( x )$
denotes the $4D$ energy-momentum tensor of the modes of the bulk field,
$\mP \approx 2.4 \times 10^{18}$ GeV
is the reduced $4D$ Planck scale
and $C^{ F F G }_{ m n q }$ is
the overlap integral of the wave functions of the $3$ modes.

We will consider only those couplings relevant for
production and decay.
Since $q \bar{q}$ annihilation to
KK graviton is Yukawa-suppressed, the production is dominated
by gluon fusion. The coupling of gluons
to KK gravitons is given by the above formula with \cite{Davoudiasl:2000wi}:
\begin{eqnarray}
C^{ A A G }_{ 0 0 n } & = &
e^{ k \pi R }
\frac{ 2 \left[ 1 - J_0 \left( x_n^G \right) \right] }{ k \pi R
\left( x_n^G \right)^2 | J_2 \left( x_n^G \right) | }
\label{G1gg}
\end{eqnarray}
where $J_q$ denote Bessel functions of order $q$ 
and (as mentioned already in the introduction) $x^G_n = 3.83, 7.02,\ldots$ gives masses of
the KK gravitons, $m^G_n = k e^{ - k \pi R } x^G_n$, while gauge KK masses
are given by
$m^A_n = k e^{ - k \pi R } x^A_n$ with $x^A_n= 2.45, 5.57,\ldots$.
We see the volume-suppression (as compared to $ 1 / \hbox{TeV}$) in the coupling, as mentioned in the introduction.
For simplicity, we neglect brane-localized kinetic terms
for both graviton and gauge fields.
Thus, we have
\begin{eqnarray}
m^G_1 & \approx 1.5 m^A_1,\label{eq:mG1}
\label{KKgKKG}
\end{eqnarray}
for the lightest KK masses for graviton and gauge fields.

Given that we are interested in the prompt decay $G_1\to \gamma_1 \gamma$, we need 
$C^{ A A G }_{ 101 }$ in Eq.~(\ref{general}), which is given in Ref.~\cite{Davoudiasl:2000wi} by the following overlap of profiles [using re-scaled
version of $z$ coordinate from Eq.~(\ref{eq:yzcoordinate}) and up to 
corrections of order $(k\pi R)^{-2} \sim 10^{-3}$] 
\beq
C^{AAG}_{101} = e^{k\pi R}\frac{2}{\sqrt{2 k\pi R}}\int_{e^{-k \pi R}}^1 dz\, z^2\, 
\frac{J_1(x^A_1 z) + \alpha^A_1 Y_1(x^A_1 z)}{|J_1(x^A_1) + \alpha^A_1 Y_1(x^A_1)|}\frac{J_2(x^G_1z)}{|J_2(x^G_1)|}\,,
\label{C101}
\eeq
where $Y_1$ is a Bessel function of order 1 and 
$\alpha^A_1 = - J_0(x^A_1)/Y_0(x^A_1)$. 
Eq.~(\ref{C101}) explicitly shows the expected $(k \pi R)^{-1/2}$ suppression.  
The tensor structure of the requisite $G_1\gamma_1\gamma$ coupling has been worked out in other extra 
dimensional settings \cite{Macesanu:2003jx,Feng:2003nr}\footnote{For important comments on typographical errors in these works, 
see our Reference section.} and would hold for our warped 5D scenario.  It turns out that the Lorentz structure of the coupling is 
the one (up to symmetry factors) of $G_1\gamma\gamma$ (see, for example, Refs.~\cite{Giudice:1998ck,Han:1998sg}) and does not depend
on KK photon mass.

Here we would like to point out the different functional dependence of the rate for our $G_1$ 
cascade decay process on $k\pi R$ compared to the conventional heavy SM final states.  As discussed before, 
warped bulk flavor models, the dominant coupling of the $G_1$ to the SM is the one to IR-brane localized fields, as schematically 
presented in Eq.~(\ref{G1tRH}), while the main $pp$ collider production mechanism is through the gluon coupling in Eq.~(\ref{G1gg}).  
Hence, the total rate for $gg\to G_1\to WW, ZZ,\ldots$ goes like $\sim (k\pi R)^{-2}$ whereas for $gg \to G_1 \to \gamma_1 \gamma$, 
the rate scales with $(k \pi R)^{-3}$, as suggested by the forms of $C^{AAG}_{001}$ and $C^{AAG}_{101}$.  Roughly speaking, we 
then expect that the ratio of the rates into heavy SM states and our final state should scale as $k \pi R$, which could be used to 
extract information on the 5D parameter $k R$.  Hence, a measurement of our process following potential discovery in the 
conventional (and likely dominant) SM modes can shed light on the underlying 5D parameters of the warped model.  

\section{\label{sec:signal} Signal, Background, and Cuts}

\label{cuts}

We now address the calculation of the rate for our process $pp \to G_1 \to \gamma_1 \gamma$, with 
$\gamma_1 \to W W \to (jj) \ell \nu$ (see FIG.~\ref{fig:decay}) in the framework outlined above.  The couplings of the KK graviton  were discussed 
before.  To implement a full simulation, we also adopted the model of Ref.~\cite{Agashe:2007ki}, where the couplings of the 
KK photon are given.  Here, the other details of the model are not very crucial for our discussion, since 
we are focused on the photon and its KK mode which does not mix with any other state due to gauge invariance and 
hence our results have less model dependence.  We adopt the values of parameters in Ref.~\cite{Agashe:2007ki}, which 
would yield $\textnormal{BR}(\gamma_1\rightarrow W^+W^-) \simeq 0.44$, as a typical value. This can be obtained from the 
approximate formulas \cite{Agashe:2007ki}  
\bea
\Gamma(\gamma_1\rightarrow t \bar{t}) &\approx& N_c Q_t^2 \frac{e^2(\kappa_L^2+\kappa_R^2)m_1^{\gamma}}{24\pi} \\
\Gamma(\gamma_1\rightarrow W_L^+W_L^-) &\approx& \frac{\left(e \;\sqrt{k\pi R}\right)^2 m_1^{\gamma}}{48\pi}\, 
\eea 
where $Q_t$ is the electric charge of top quark, and the $\kappa_L$ and $\kappa_R$ couplings are given by
\bea
\kappa_L &=& -\frac{1.13}{\sqrt{k\pi R}}+0.2\sqrt{k \pi R} \\
\kappa_R &=& -\frac{1.13}{\sqrt{k\pi R}}+0.7\sqrt{k \pi R}.
\eea
\subsection{Kinematic Features of the Signal and Cuts}

\begin{figure}[t]
\centering
\includegraphics[scale=0.6]{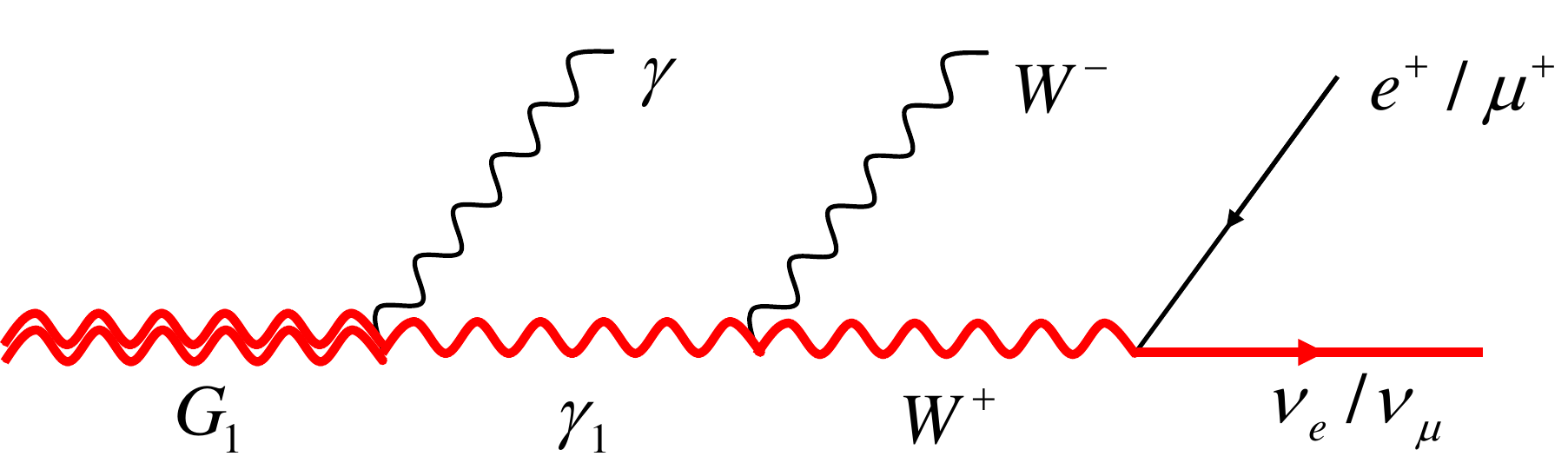}
\caption{\label{fig:decay} Decay topology of the signal process.}
\end{figure}

The signal process at hand is characterized by a sequential cascade decay of a heavy resonance $G_1$ diagrammed in FIG.~\ref{fig:decay} where the hadronic $W$ is considered as fully visible.
Since all visible particles are completely distinguishable, many of the distinctive kinematic features can be easily 
utilized without suffering from any possible combinatorial issues.  Our signal includes a very energetic photon, due to 
the high mass of the parent resonance $G_1$, $m^G_1 \gsim 1$~TeV, which sets the energy scale of our process.  
The energy of the emitted photon in the rest frame of the KK graviton is 
\bea
E_{\gamma}^{\textnormal{rest}}=\frac{\left( x_1^G \right)^2-\left( x_1^{\gamma} \right)^2 }{ x_1^G x_1^{\gamma} }\frac{m_1^{\gamma}}{2} \approx \frac{m_1^{\gamma}}{2}.
\eea 
Since most KK gravitons are produced nearly at rest, the above relation indicates that we can typically expect a very hard $\gamma$ in our signal.  Given the 
approximate relationship $m^G_1\sim 1.5 m^A_1$, we also expect $m^\gamma_1 \gsim 1$~TeV.  

We note that $m^\gamma_1$ can be constrained 
by the existing direct lower bound on the lightest KK gluon mass which is about 2.5~TeV~\cite{Chatrchyan:2013lca,TheATLAScollaboration:2013kha}.  
However, these bounds have some model dependence, where 
certain strength of coupling between light quarks (initial states at LHC) and the KK gluon is assumed.  In any event, even for $m^\gamma_1 \gsim 1$ TeV, we expect 
that each $W$ to have a momentum $P_W \gsim 0.5$~TeV.  Hence, the expected opening angle for the jets in $W\to jj$ is roughly $2 m_W/P_W\lsim  0.3$, 
which is below the typical jet cone radius $R_\text{jet}=0.4$ assumed in typical analyses.  We will then assume that our hadronically 
decaying $W$ in the final state cannot be reconstructed with the usual techniques and will appear as a single jet in the detector.  
We will further assume that jet-shape techniques can be applied to tag merged two-prong $W$-jets and reject the single-prong QCD jet background, as we will discuss later.      
Considering the typical mass scale of the KK particles in our signal process, it is fairly reasonable to set the minimum transverse momentum 
$P_T$ of jets and leptons, including the missing neutrino momentum, to be $\mathcal{O}(100$ GeV$)$. 

There are two main invariant masses defined by KK graviton and KK photon involved in our signal process. The associated invariant mass window cuts can play a key role in 
rejecting background events. In order to reconstruct those two invariant masses, however, we first obtain the four momentum of the invisible neutrino. Its two transverse momentum components can be easily reconstructed by the missing transverse momentum constraint. The energy and the longitudinal momentum component can be obtained by solving the neutrino and $W$ mass-shell equations. As is well-known, relevant solutions have a two-fold ambiguity. Therefore, for a given event, if either of the solutions satisfies the two invariant mass window cuts, then it is regarded as accepted in our analysis.
\begin{figure}[t]
\centering
\includegraphics[width=7.6cm]{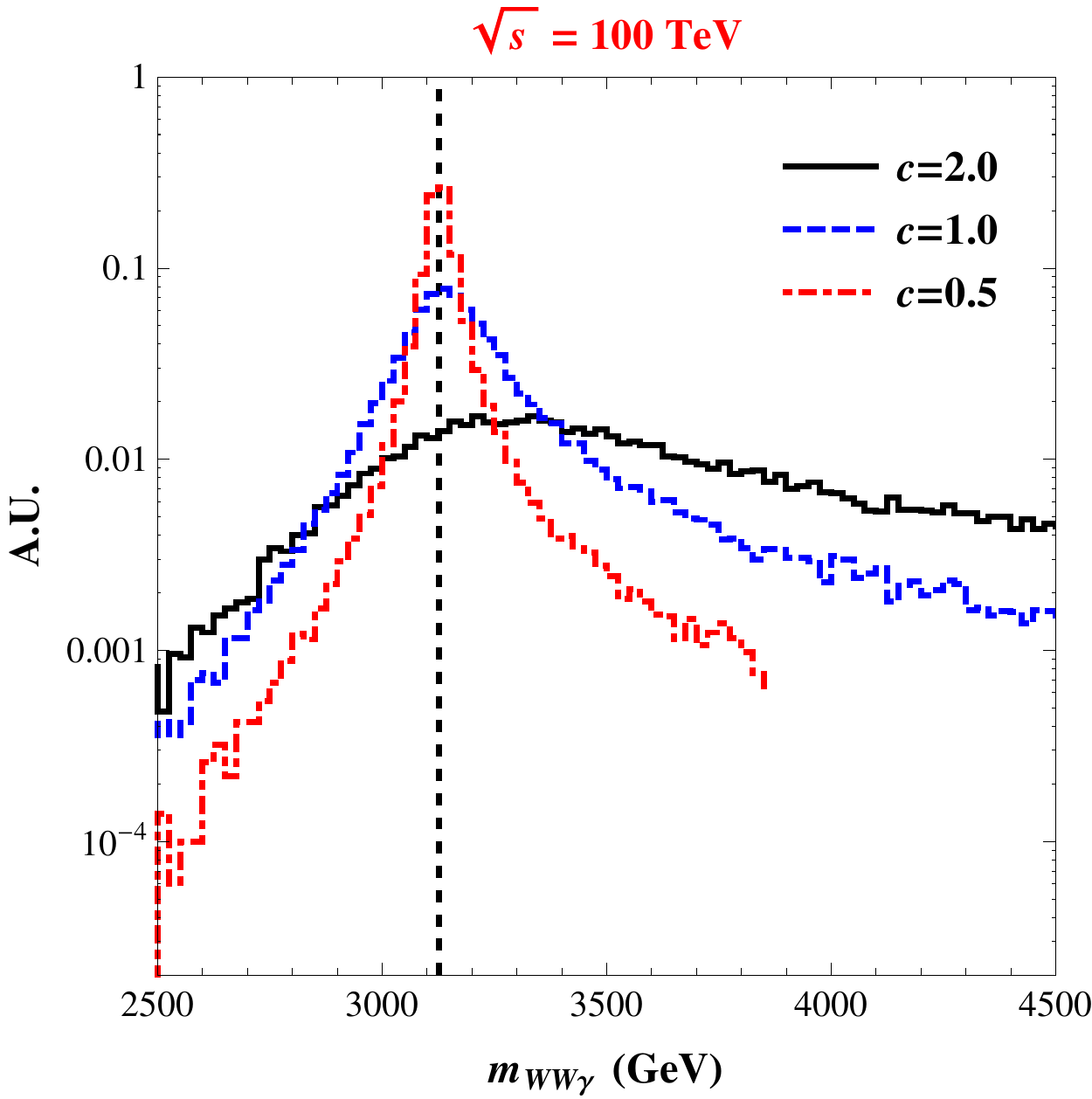}
\includegraphics[width=7.3cm]{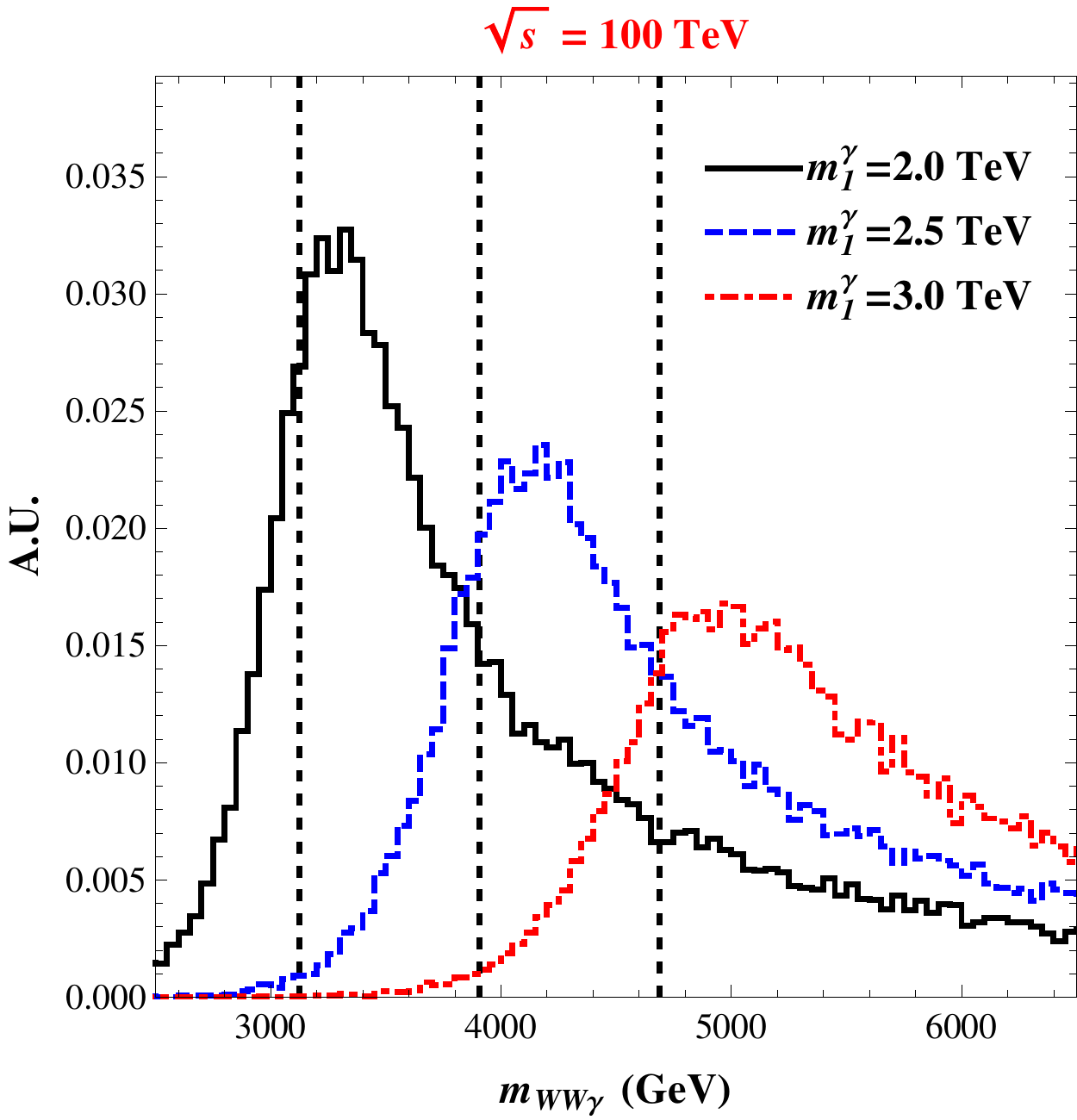}
\caption{\label{fig:decaywidthG1} Unit-normalized $m_{WW\gamma}$ distributions. The left panel shows such distributions for three different $c$ parameters (2.0, 1.0, and 0.5) with $m_1^{\gamma}$ fixed to $2$ TeV, whereas the right panel shows them for three different masses of KK photon (2, 2.5, and 3 TeV) with $c$ parameter fixed to 2.0.  
Here, $\sqrt{s}=100$~TeV for both panels. The dashed lines indicate the relevant theory expectation.}
\end{figure}
\begin{figure}[t]
\centering
\includegraphics[width=7.6cm]{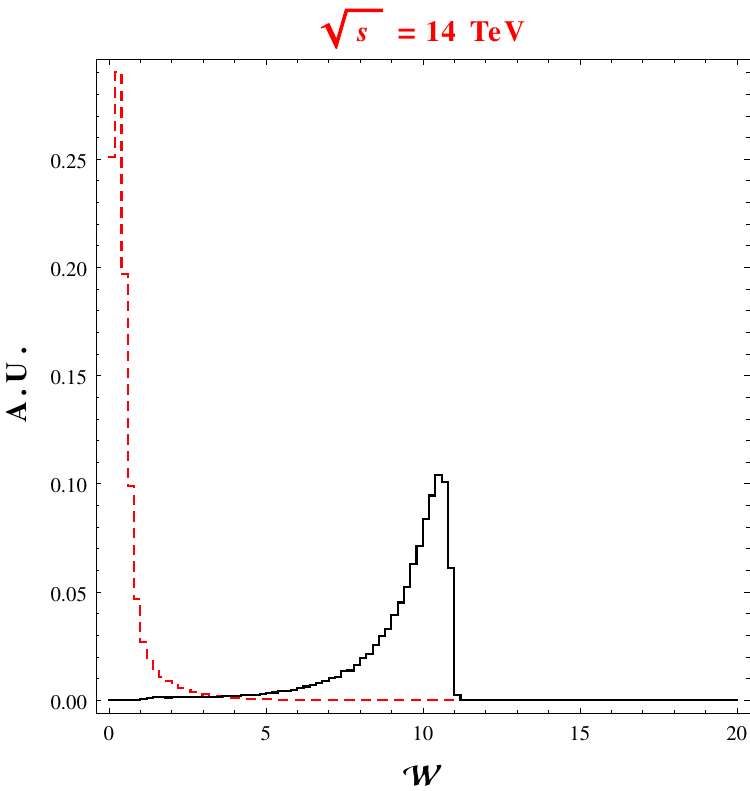}
\includegraphics[width=7.6cm]{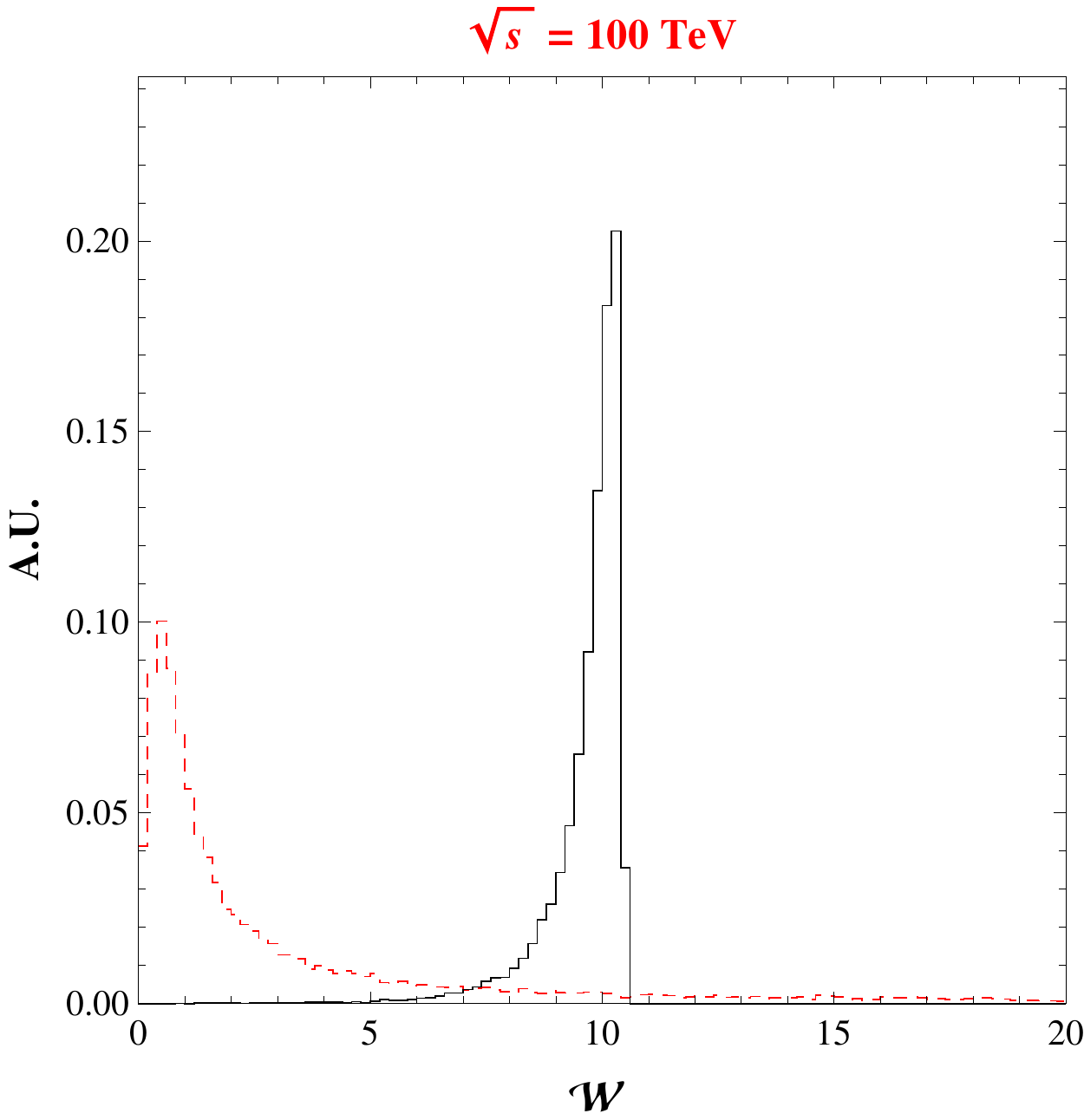}
\caption{\label{fig:14W} Unit-normalized distributions of the ${\cal W}$ variable for the dominant background $Wj\gamma$ (black solid histogram) and the signal (red dashed histogram) events. The left (right) panel takes Study Point 1 (3) at $\sqrt{s}=14$ TeV ($\sqrt{s}=100$ TeV) in Table~\ref{tab:cutflow14TeV} (Table~\ref{tab:cutflow100TeV}) as the relevant mass parameter. The distributions are plotted with the events passing only basic cuts defined in Section~\ref{sec:results}. 
}
\end{figure}
Once both of the solutions are obtained, we compute $m_{\gamma \ell jj \nu}^{(i)}$ and $m_{\ell jj \nu}^{(i)}$ ($i=1,2$) to see if either would 
meet the criteria of the invariant mass window, for example, 
\bea
m_1^{\gamma}-\Gamma_1^{\gamma}< &m_{\ell jj \nu}& < m_1^{\gamma}+\Gamma_1^{\gamma}  \label{eq:masswindowkka}\\
m_1^{G}-\Gamma_1^{G}< &m_{\gamma \ell jj \nu}& < m_1^G+\Gamma_1^G\label{eq:masswindowkkg}
\eea
where $\Gamma_1^G$ and $\Gamma_1^{\gamma}$ are the widths of $G_1$ and $\gamma_1$, respectively. 
However, if the $c$ parameter is large enough, e.g., $c=2$, the second criterion would {\it not} keep as many signal events as possible. More quantitatively, 
from Eq.~(\ref{eq:G1decawidth}) we see that for large values of the $c$ parameter, the KK graviton width becomes larger than $\sim 20\%$ of its mass
and the corresponding Breit-Wigner distribution could get skewed from the expectation. 
This is because the gluon-gluon-KK graviton coupling emerges from a dim-5 operator (in turn, due to its spin-2 nature) which would like to grow with energy, and this effect is particularly obvious when the width of the graviton is large, e.g., $c=2$.  At a 100 TeV collider, the KK graviton mass of a few TeV is in the regime of low $x$, so that we expect that the associated invariant mass distribution is not significantly affected by the gluon parton distribution function.  However, at the 14 TeV of LHC with $m^G_1$ $\sim  2$ TeV, this effect becomes smaller due to the competition between the growing gluon-gluon-KK graviton coupling and the rapid falling behavior in the regime of large $x$ in the gluon parton distribution function.
This expectation is clearly verified in FIG.~\ref{fig:decaywidthG1}. In the left panel, we vary the $c$ parameter with the mass of KK photon fixed to 2 TeV. We see that the peak position is consistent with the relevant expectation for $c=0.5,\;1$, whereas that for $c=2$ is shifted to the right.  The right panel demonstrates such a shift for three different KK photon masses, $i.e., 2, 2.5$, and 3 TeV with $c$ parameter fixed to 2. We observe that the peak position is shifted by about a half of $\Gamma_1^G$ for all cases. Based upon these observations, in the case of $c=2$ and $\sqrt{s}=100$ TeV we modify Eq.~(\ref{eq:masswindowkkg}) into an asymmetric form as
\bea
 m_1^{G}-\Gamma_1^{G}< m_{\gamma \ell jj \nu} < m_1^G+2\Gamma_1^G, \label{eq:modifiedmasswindowkkg}
\eea
which enables us to secure more signal statistics (at the risk of including more background events)
\footnote{Although such a skewness is less pronounced at the LHC14 as mentioned before, we still apply this asymmetric form to the relevant analysis later.}.

When it comes to the selection process in regard to $m_{\ell jj \nu}$ and $m_{\gamma \ell jj \nu}$, a simple way is to reject the events that fall outside the 
mass windows defined by Eqs.~(\ref{eq:masswindowkka}) and~(\ref{eq:masswindowkkg}) (or Eq.~(\ref{eq:modifiedmasswindowkkg}) for $c=2$), simultaneously. This would exclude the events where one of the invariant mass windows is marginally dissatisfied while the other is satisfied.  Given the underlying dynamics, one may expect that such a situation arises rather rarely for backgrounds.  Hence, in order to address such cases, we introduce a new {\it weighted} measure ${\cal W}$ which, by construction, considers both invariant mass windows collectively:
\bea
{\cal W}=
\frac{|m_{\ell jj \nu}-m_1^{\gamma}|}{\Gamma_1^{\gamma}}+\frac{|m_{\gamma \ell jj \nu}-(m_1^G+0.5\Gamma_1^G)|}{1.5\Gamma_1^G}.\; 
\eea
The above form roughly captures the expected features of the signal and will be used in our analyses.
FIG.~\ref{fig:14W} exhibits the ${\cal W}$ dependence for Study Points 1 and 3 in Table~\ref{tab:cutflow14TeV} and~\ref{tab:cutflow100TeV} together with the dominant background $Wj\gamma$. One can easily see a clear separation between the signal and the background. The signal events typically have very small values of ${\cal W}$, whereas the background events peak at a larger value of ${\cal W}$. This is because it is rather difficult for the reconstructed KK graviton and photon masses in background events to be within the associated invariant mass windows simultaneously. In fact, this feature is expected to be even more beneficial with smaller $c$ parameters due to the narrowness of the KK graviton mass window. Given an event, we therefore keep it if its ${\cal W}$ is smaller than a certain cut (e.g., 1) depending on the scanning point in $m_1^{\gamma}$, and otherwise, we reject it. 

Another kinematic variable is the pseudo-rapidity $\eta$ which is a good discriminator between the signal and the background events, as we generally expect that the former are 
more central than the latter.  A related variable is $\Delta R_{jj}\equiv \sqrt{(\Delta \phi^j)^2+ (\Delta \eta^j)^2}$, with $\phi$ the azimuthal angle, that defines the distance between the two jets.   As discussed above, the $W$ gauge bosons are anticipated to be highly boosted because KK photon is quite heavy. Thus we expect that the two jets coming from the decay of a $W$ are likely to be highly collimated and, in turn, merged into a single jet.  In order to capture this effect, we will require $\Delta R_{jj}<0.4$ for the selected signal events.   

\subsection{Background}

As described in the previous section, the signal process of interest is characterized by $\gamma \ell jj+\misse$. Given this collider signature, several SM processes can be considered as possible backgrounds.  The irreducible SM background is from $WW\gamma$ production.  As explained before, the kinematics of the signal 
includes merged $jj$ events from highly boosted $W$'s.  Hence, 
we expect that the most important reducible background to come from $Wj\gamma$, which includes a QCD jet.  This reducible background dominates 
over the irreducible one from $WW\gamma$ final states.  

With a photon fake rate of $\sim 10^{-4}$~\cite{Aad:2009wy}, backgrounds, such as $WWj$, with the jet faking a high energy photon are quite suppressed.  
The next type of background is $ZZ\gamma$ in which the $Z$ gauge bosons decay semi-leptonically.  Since the signal process of interest involves only a single lepton in the final state, one of the leptons from $Z$ decay should be lost. This can happen if the lepton has a transverse momentum $P_T^{\ell}$ that  is too soft to pass the relevant acceptance or it comes with a large $\eta$ which is not covered by the electromagnetic calorimeter and the muon chamber. The background events in the former case can be easily rejected by a sizable $\misse$ cut. Our simulation study suggests that the relevant event rate should be small enough to neglect the $ZZ\gamma$ contribution.   
Another potential reducible background is $Wjj\gamma$ in which again the two jets are produced via QCD.  However, since 
we are assuming a signal selection criterion $\Delta R_{jj}<0.4$ in our analysis, this background can be assumed to be rejected as a dijet final state.  

\section{Results/Discovery Reach \label{sec:results}}

\label{results}

\begin{figure}[t]
\centering
\includegraphics{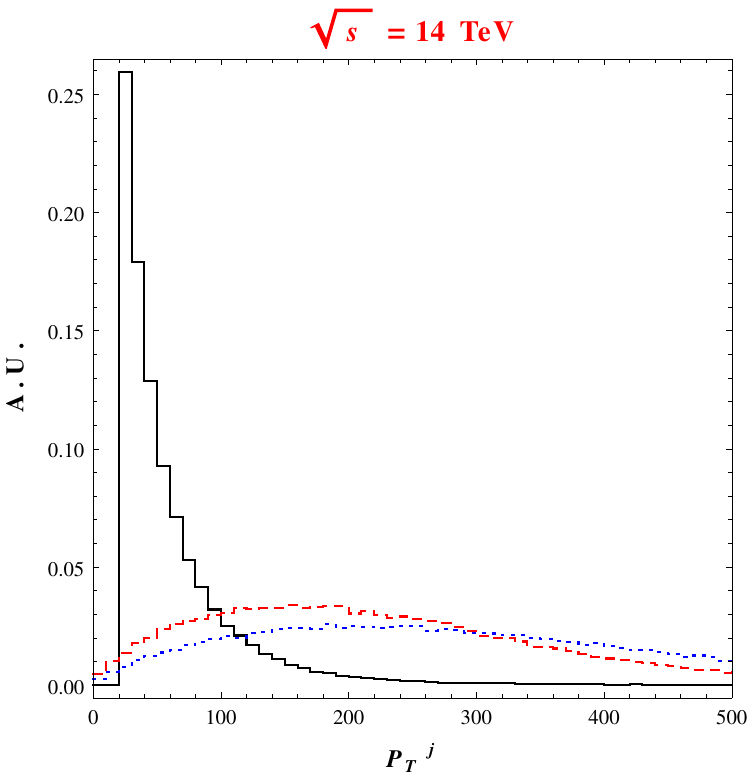}
\includegraphics[width=7.3cm]{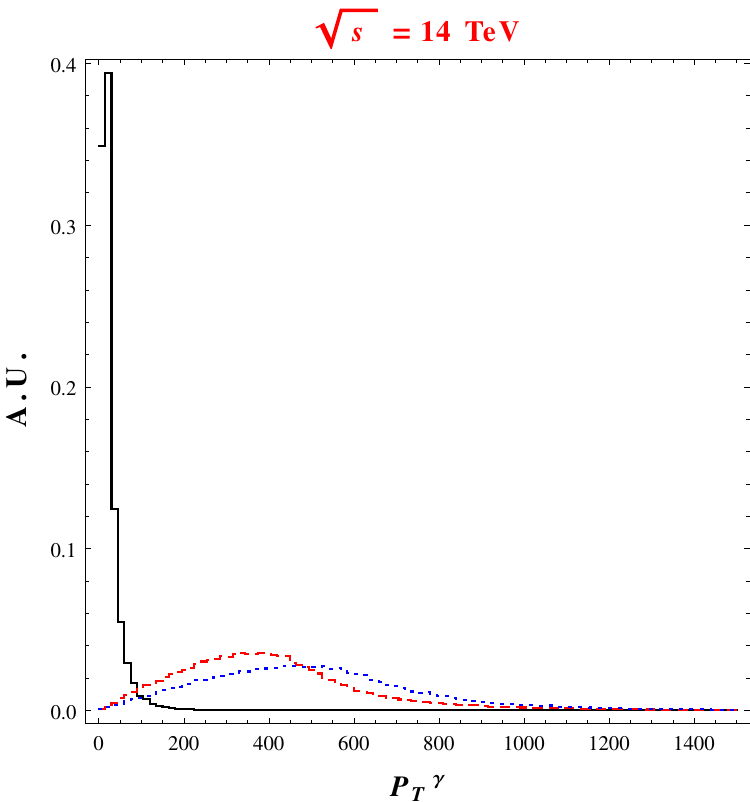}
\caption{\label{fig:14pt} Unit-normalized distributions of $P_T^j$ in the left panel and $P_T^\gamma$ in the right panel for the dominant background $Wj\gamma$ (black solid) and two study points, $m_1^G=1.5$ TeV (red dashed) and $m_1^G=2$ TeV (blue dotted) with $c=2$.}
\end{figure}

\begin{figure}[t]
\centering
\includegraphics{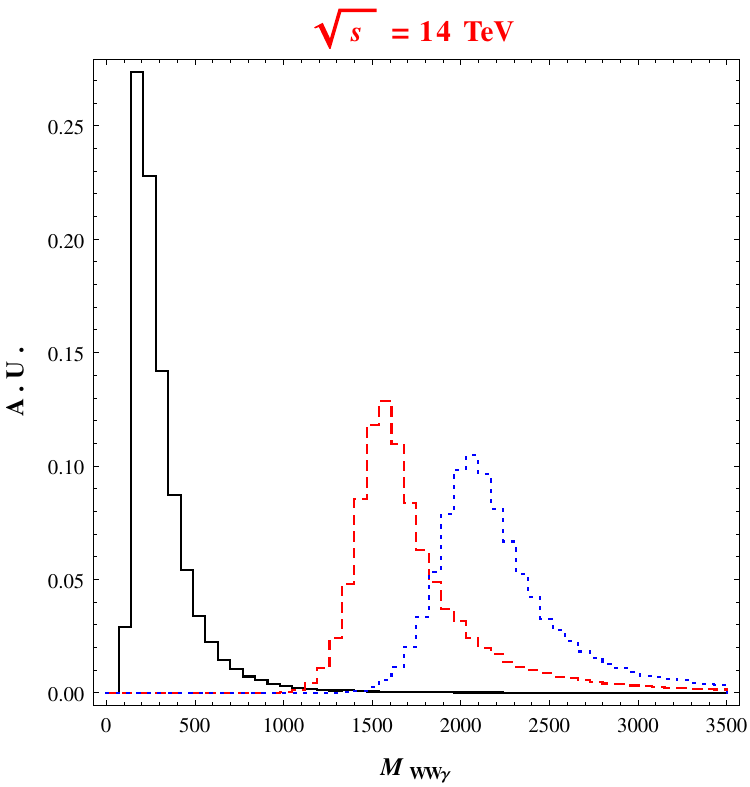}
\includegraphics{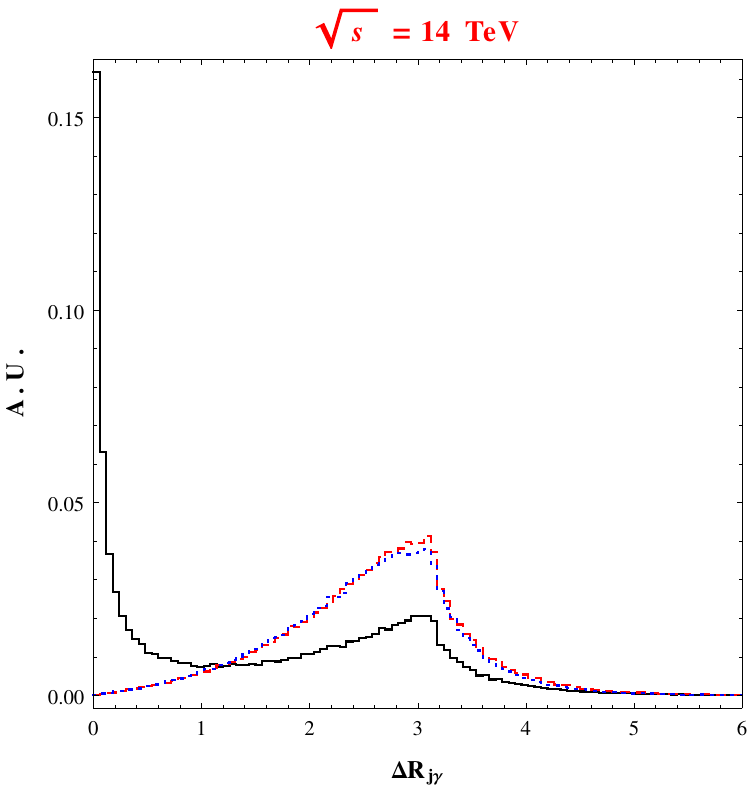}
\caption{\label{fig:14drja} Unit-normalized distributions of the invariant mass of the two $W$ bosons and photon in the final states $M_{WW\gamma}$ in the left panel and $\Delta R_{j\gamma}$ in the right panel for the dominant background $Wj\gamma$ (black solid) and two study points, $m_1^G=1.5$ TeV (red dashed) and  $m_1^G=2$ TeV (blue dotted) with $c=2$.}
\end{figure}

In this section, we discuss the discovery opportunity of the KK graviton for the study points (SPs) based upon the cuts given in Tables~\ref{tab:cutflow14TeV} and~\ref{tab:cutflow100TeV}. In Table~\ref{tab:cutflow14TeV} we study two points, SP1 and SP2 at $\sqrt{s}=14$ TeV, which have $m_1^G=1.5$ TeV and 2 TeV with $c=2$, respectively.
We also show two study points, SP3 and SP4 at $\sqrt{s}=100$ TeV, which have $m_1^\gamma=2.5$ TeV and 3 TeV with $c=2$, respectively.
The value of $c$ here is near the maximum for a valid perturbative description~\cite{Agashe:2007zd} 
and would hence  give us a rough estimate of largest expected reach.
As mentioned in the previous section, the dominant SM background is $Wj\gamma$, and
we will discuss how to reduce this background by applying a series of cuts.
First of all, FIG.~\ref{fig:14pt} shows the distributions of the transverse momenta of the leading jet $(P_T^j)$ and the photon $(P_T^\gamma)$ for the
background and signals. For both distributions,
the background peaks in the low $P_T$ region while the signal events tend to have larger values of $P_T$ above 100 GeV.
As a result, one can easily see that $P_T^j > 100$ GeV and $P_T^\gamma > 100$ GeV are very powerful cuts to reduce this background, which confirms our expectation in the previous section. 
Furthermore, since the KK graviton is almost produced at rest, its decay products tend to decay back to back, which implies that the leading
jet and photon also have tendency to be back to back, as shown in FIG.~\ref{fig:14drja}. However, the background events are mainly produced through QCD processes and do not have this signature. Hence, it is very useful to require the separation between the leading jet and photon $(\Delta R_{j\gamma})$ to be larger than $\sim$ 2.
Finally, the invariant mass formed by the two $W$ gauge bosons and the photon can be a powerful signal-background discriminator, and this observation has been translated into the $\mathcal{W}$ measure whose performance was already confirmed in FIG.~\ref{fig:14W}. 

To conduct the Monte Carlo simulation for the signal and background processes at the parton level, we employ the matrix element generator \texttt{MG5\_aMC@NLO}~\cite{Alwall:2014hca} and CalcHEP 3.4~\cite{Belyaev:2012qa}, taking the parton distribution functions of \texttt{NNPDF23}~\cite{Ball:2012cx}. For an implementation of the warped hierarchy/flavor model, we first employ existing model files in Ref.~\cite{CP3}. The vertex structure of $G_1\gamma_1 \gamma$, which is absent in the model files, is written based on the corresponding structure encoded in $G_1\gamma\gamma$. Various decay modes of the KK photon are implemented by modifying the existing vertices in the model files.   

\begin{table}[h]
\centering
\begin{tabular}{c|c c|c c}
 & SP1 \;\;\;\;\;    & SP2  & $WW\gamma$ & $Wj\gamma$ \\
 \hline \hline
No cut                          & 0.067 & 9.30 $\times 10^{-3}$     &  -- & --\\
Basic cuts                      & 0.060 & 8.61 $\times 10^{-3}$  & (58.51) & (2.68$\times10^4$) \\
$p_T^{\gamma}>100$              & 0.592 & 8.54 $\times 10^{-3}$   & 3.56 & 6.69$\times10^2$ \\
$p_T^{j/\ell}>100$              & 0.033 & 6.08 $\times 10^{-3}$   & 0.18 & 92.25 \\
$|\eta^{\textnormal{all}}|<2.0$ & 0.029 & 5.44 $\times 10^{-3}$   & 0.15 & 71.81 \\
$E_T^{\textnormal{miss}}>100$   & 0.023 & 4.86 $\times 10^{-3}$   & 0.07 & 19.18\\
$\Delta R_{j\gamma}>2$          & 0.017 & 3.52 $\times 10^{-3}$  &  0.03 & 6.93\\
$\Delta R_{jj}<0.4$             & 0.014 & 3.30 $\times 10^{-3}$   & 0.02 & -- \\
$60<m_{jj}<100$                 & 0.014 & 3.26 $\times 10^{-3}$  &  0.02 & -- \\
\hline
${\cal W}<0.5$(SP1)                    & 0.009 & --                      &   0 & 0.056  \\
${\cal W}<0.9$(SP2)                    & --    & 2.69 $\times 10^{-3}$    &  0 & 0.047 \\
\hline \hline
$\mathcal{L}$ (ab$^{-1}$)       & 3 & 3       & 3   & 3 \\
\hline
Number of events (SP1)           & 27    &--     & 0 & 168\\
Number of events (SP2)           & --      & 8 &   0 &141    \\
\hline
$S/\sqrt{B}$                    & $3.2\sigma$ & $1.1\sigma$ & -- & --
\end{tabular}
\caption{\label{tab:cutflow14TeV} Signal and background cross sections in fb in the sequence of cuts for a study point --SP1: $m_1^{G}=1.5$ TeV with $c=2$, and SP2: $m_1^{G}=2$ TeV with $c=2$, and dominant backgrounds at a $pp$ collider of $\sqrt{s}=14$ TeV. Background cross sections at leading order in parentheses were evaluated with the basic cuts such as $p_T^j>20$ GeV, $p_T^{\gamma/\ell}>10$ GeV, $|\eta^j|<5$, $|\eta^{\gamma/\ell}|<2.5$, and $\Delta R_{jj/j\gamma}>0.01$ to avoid any possible divergence. All momenta and masses are in GeV.}
\end{table}

\begin{table}[h]
\centering
\begin{tabular}{c|c  c|c c}
 & SP3 & SP4 & $WW\gamma$ & $Wj\gamma$ \\
 \hline \hline
No cut                          & 0.4    & 0.13 &  --   & --\\
Basic cuts                       & 0.35   & 0.12 & (391) & (1.68$\times10^5$) \\
$p_T^{\gamma}>600$              & 0.31   & 0.11 & 1.81  & 132.0 \\
$p_T^{j/\ell}>150$               & 0.26   & 0.10 & 0.28  & 42.5 \\
$|\eta^{\textnormal{all}}|<2.0$  & 0.21   & 0.08 & 0.19  & 29.6 \\
$E_T^{\textnormal{miss}}>150$    & 0.20   & 0.077 & 0.10  & 13.1\\
$\Delta R_{jj}<0.4$              & 0.19   & 0.077 & 0.09  & -- \\
$60<m_{jj}<100$                  & 0.19   & 0.077 & 0.09  & -- \\
\hline
${\cal W}<0.9$(SP3)                    & 0.03    &--     & 0.0025& 0.29\\
${\cal W}<2.0$(SP4)                   & --      & 0.014 &  0.0055&1.19    \\
\hline \hline
$\mathcal{L}$ (ab$^{-1}$)       & 3     &  3     & 3   & 3 \\
\hline
Number of events (SP3)           & 90    &--     & 7.5  & 870\\
Number of events (SP4)           & --      & 42  &  16.5 &3570    \\
\hline
$S/\sqrt{B}$                    & $5.0\sigma$  & $1.1\sigma$ & -- & --
\end{tabular}
\caption{\label{tab:cutflow100TeV} Signal and background cross sections in fb in the sequence of cuts for a study point --SP3: $m_1^{\gamma}=2.5$ TeV with $c=2$, and SP4: $m_1^{\gamma}=3$ TeV with $c=2$ and dominant backgrounds at a $pp$ collider of $\sqrt{s}=100$ TeV. Background cross sections at leading order in parentheses were evaluated with the basic cuts such as $p_T^j>20$ GeV, $p_T^{\gamma/\ell}>10$ GeV, $|\eta^j|<5$, $|\eta^{\gamma/\ell}|<2.5$, and $\Delta R_{jj/j\gamma}>0.01$ to avoid any possible divergence. All momenta and masses are in GeV.}
\end{table}
The total number of signal (denoted by $S$) and background (denoted by $B$) events are obtained by using the following formulae:
\begin{eqnarray}
 S&=&\epsilon_W \times N_S, \nonumber \\
 B&=&\epsilon_W \times N_{WW\gamma}+2\times (1-\epsilon_j)\times N_{Wj\gamma},\nonumber
\end{eqnarray}
where $N_{WW\gamma}$ and $N_{Wj\gamma}$ are number of events for the
$WW\gamma$ and $Wj\gamma$ backgrounds. $\epsilon_W=0.5$ is the tagging rate for a $W$-jet and $\epsilon_j=0.95$ is the rejection rate
for a QCD jet \cite{CMS:2014joa}. To be conservative we include a factor of two for the $Wj\gamma$ background to account for the next to
leading order corrections.

We find that our signal can be detected at the $3\sigma$ level for a warped graviton up to a mass of order 1.5~TeV, with an integrated luminosity of 3~ab$^{-1}$.
This value of $G_1$ mass is well below those implied by precision data~\cite{Carena:2006bn,Carena:2007ua} and the direct bounds on KK gluon masses~\cite{Chatrchyan:2013lca,TheATLAScollaboration:2013kha}, both of which
roughly yield $m^G_1\gsim 4$~TeV.  However, experimental results from ATLAS and CMS \cite{Khachatryan:2014gha,ATLAS:039} are not yet at a level
that would rule out $m^G_1\sim 1.5$~TeV {\em directly}.  
Bounds from precision data are not entirely rigid and depend on what
is considered, e.g., the {\em natural} parameter space of the theory and their degree of tuning (see, for example,
Ref.~\cite{Delaunay:2010dw}).  

As for the KK gluon bounds, they do depend
on the profiles of the light quarks, which are the initial states in KK gluon production.  If, for example, some of the light quarks
have a flat profile, they would decouple from the KK gluon, which would degrade the direct KK gluon mass bound,
but then they would contribute, albeit at a volume-suppressed level,
to KK graviton production, hence enhancing both the direct bounds and also the predicted reach for our channel.
Another example is when the KK gluon is allowed to decay into other new heavy particles, the bound on the KK gluon (and hence
the indirect one on KK graviton) can be
significantly reduced~\cite{Greco:2014aza}.
A definite statement about how the various bounds can be affected by modulation of quark profiles and other parameters
requires a detailed study that is not within the scope of our work. However, the above considerations highlight how the expectations
for the mass of $G_1$ can be affected by varying the underlying model parameters.

The cuts presented in Table~\ref{tab:cutflow100TeV} reflect the enhanced capabilities of a 100 TeV hadron collider, where
the kinematics are characterized by larger energy scales.  Here, we see that $m^G_1 \sim 3.7$~TeV can allow
a $5\sigma$ detection of our cascade decay signal, given an integrated luminosity of 3~ab$^{-1}$.  This mass scale for $G_1$
is both roughly consistent with precision electroweak constraints, as well as the implied current bounds from non-observation of KK gluons.


\section{Conclusions and Outlook}

\label{conclude}

A distinct signature of warped models of hierarchy and flavor models is the appearance of a TeV-scale spin-2 resonance, the lightest KK excitation of the graviton $G_1$. Conventional search strategies for $G_1$ focus on heavy SM final states, such as $t\bar t$ or longitudinal $ZZ$, that have sizeable $G_1$ couplings, but often require reconstruction with various degrees of efficiency or suppression due to the small branching fractions.

In this work, we considered a novel ``cascade decay" mode of $G_1$ into an on-shell KK photon $\gamma_1$ and a massless photon, $G_1 \to \gamma_1 \gamma$, with the subsequent decay $\gamma_1 \to W W \to jj \ell \nu$. The high energy photon provides a distinct and elementary final state signal which can be detected efficiently. While the photon coupling is suppressed by a 5D ``volume factor," the heavy $\gamma_1$ has strong coupling to $G_1$ which makes this channel only semi-suppressed.  Consequently, the dependence of the rate of our channel on the underlying 5D volume factor is distinct from those of the conventional heavy SM pair decay channels. This implies that measurements of $G_1$ in our cascade mode in conjunction with conventional final states could in principle yield new handles on the 5D underlying theory. We found that the high luminosity run of the 14 TeV LHC with an integrated luminosity of 3~ab$^{-1}$ can find $3\sigma$ evidence for $G_1 \to \gamma_1 \gamma$, for $G_1$ at 1.5~TeV. While this mass scale for $G_1$ is below that expected from precision electroweak and flavor data, or the lack of evidence for the warped KK gluon states below $\sim 2.5$~TeV, it is not constrained by direct searches for $G_1$.  Precision constraints and the KK gluon bounds, though well-motivated, might be subject to various model dependent assumptions, such as fermion profiles in the 5D bulk:
simple modifications of the minimal model might then allow lower KK masses.
On the other hand, the profiles of the gauge and graviton modes are largely determined by the underlying background geometry, that is its ``5D volume," and are hence characterized by the basic properties of warped models.
Thus, couplings of 
$G_1$
to gluons in initial state and to
 $\gamma_1$ and $\gamma$ in final state, for the process that we studied here, are relatively speaking fixed.
While the {\em total}
decay width, which also enters the calculation of the above signal, is subject to additional assumptions, including presence of 
extra light states, our eventual idea anyway was to {\em combine} this channel with the ones studied earlier 
in order to measure this (in general) independent parameter. 

Looking to the future, we also examined the prospects for detecting our signal at a 100 TeV hadron collider.  Here, as expected, the reach markedly improves and an integrated luminosity of 3~ab$^{-1}$ can yield $5\sigma$ evidence for $G_1 \to \gamma_1 \gamma$ up to roughly 4~TeV. Such masses for $G_1$ can be consistent with precision data requirements, as well as the current bounds on KK gluon mass.  One could envision the detection of $G_1$ in more conventional modes that have larger rates, followed by a more detailed study of our or other cascade channels that probe various bulk interactions and geometric properties of the underlying warped model.

Finally, we would like to point out an interesting feature that we observed during our analysis. Namely, we found that the 
invariant mass distribution of final states from KK graviton production is skewed from the expectation of the usual Breit-Wigner shape.
The underlying reason is that the 
couplings of the KK graviton grow with energy, in turn, because of its spin-2 nature (i.e., this happens at the leading order).
Note that this consideration applies at the parton-level and competes with the opposite tendency of the PDF's to grow towards lower invariant masses.
Of course, such a feature arises, in general, for a decay via a
higher-dimensional operator.
Thus, it 
might be possible to utilize this behavior in order to 
distinguish a KK graviton resonance from spin-1 or 0, where 
%
%
often the 
%
%
dominant coupling is constant with energy.

%



\section*{Acknowledgments}
We thank Sally Dawson, Andrew Larkoski, Ian Lewis, Konstantin Matchev, Myeonghun Park, Brock Tweedie, and Cen Zhang for useful discussions. D.~K. also thanks Asia Pacific Center for Theoretical Physics in Pohang, South Korea for hospitality during the writing of part of this paper. 
K.~A. is supported in part by NSF Grant No.\ PHY-1315155 and the Maryland Center for Fundamental Physics.
The work of C.-Y.~C. and H.~D. is supported in part by the US DOE Grant DE-AC02-98CH10886. D.~K. was supported in part by NSF Grant No. PHY-0652363, and also acknowledges the support from the LHC Theory Initiative postdoctoral fellowship (NSF Grant No. PHY-0969510).  We thank the authors of Ref.~\cite{Feng:2003nr} for communication
about the correct form of $G_1\gamma_1\gamma$ coupling in their paper.



\end{document}